\documentclass[a4paper,11pt]{article}

\usepackage{graphicx}  
\usepackage{dcolumn}   
\usepackage{bm,relsize}        
\usepackage{amssymb, amsmath}
\usepackage{textcomp}
\usepackage{wasysym}
\usepackage{slashed}
\usepackage{caption, subcaption}
\usepackage{multirow}
\usepackage{gensymb}
\usepackage{subcaption}
\usepackage{colortbl}
\usepackage{booktabs}
\usepackage{tabularx}
\definecolor{headergray}{gray}{0.9}
\definecolor{rowgray}{gray}{0.97}

\usepackage[nolist, nohyperlinks]{acronym}

\usepackage{lipsum, color}
\usepackage[dvipsnames,svgnames,table,x11names]{xcolor}
\usepackage{braket}
\usepackage{jheppub} 
\usepackage{booktabs}
\usepackage{pdflscape}
\usepackage[utf8]{inputenc} 

\usepackage{booktabs}
\usepackage{tabularx}

\usepackage{mathrsfs}
\usepackage{bm,amssymb,slashed,graphicx,multirow,soul,mathtools,xspace,array,tikz,amsmath, gensymb}
\usepackage{siunitx} 
\usepackage{float}   
\usepackage{cancel}
\allowdisplaybreaks
\usepackage{ bbold }
\usepackage{caption,subcaption}
\usepackage{hyperref}
\usepackage{colortbl}
\usepackage{tcolorbox}

\usepackage{comment}

\usepackage[capitalise, english]{cleveref}

\usepackage{mathrsfs}

\definecolor{nicered}{rgb}{0.7,0.1,0.1}
\definecolor{nicegreen}{rgb}{0.1,0.5,0.1}
\definecolor{violet}{rgb}{0.7,0.3,0.3}
\hypersetup{colorlinks,citecolor= nicegreen,linkcolor= nicered}

\newcommand{\be}{\begin{equation}}
\newcommand{\ee}{\end{equation}}

\newcommand{\dd}[0]{\mathrm{d}}

\newcommand{\beq}{\begin{equation} }
\newcommand{\eeq}{\end{equation}} 
\newcommand{\bi}{\begin{itemize} }
\newcommand{\ei}{\end{itemize} }

\definecolor{Red}{rgb}{1.,0.,0.}
\definecolor{Grn}{rgb}{0.,0.75,0.}
\definecolor{Blu}{rgb}{0.,0.,1.}
\definecolor{Pink}{rgb}{1,0.08,0.58}

\usepackage[T1]{fontenc} 

\usepackage{amsmath,amssymb,epsfig,color,slashed}
\allowdisplaybreaks  

\newcommand{\Op}{\mathcal{O}}

\begin{document} 


\title{Inferring dark matter masses and interactions from high recoil energy events in LUX-ZEPLIN}

\author[a]{Stefano Palmisano,}
\author[b]{Michele Tammaro,}
\author[a]{Andrea Tesi}
\affiliation[a]{INFN Sezione di Firenze, Via G. Sansone 1, I-50019 Sesto Fiorentino, Italy}
\affiliation[b]{Galileo Galilei Institute for Theoretical Physics, Largo Enrico Fermi 2, I-50125 Firenze, Italy}

\emailAdd{stefano.palmisano@fi.infn.it}
\emailAdd{michele.tammaro@fi.infn.it}
\emailAdd{andrea.tesi@fi.infn.it}

\date{\today}

\preprint{}

\abstract{LUX-ZEPLIN has recently revealed a single nuclear recoil event compatible with a recoil energy of $E_R\simeq248\,\mathrm{keV}$, in a region with a small expected background. Interpreting this as a signature of dark matter, we compute the posterior probability distribution of the parameters of the dark matter scattering off xenon nuclei. We retain the exact two-body kinematics for an arbitrary mass splitting between the incoming and outgoing dark-sector states, thus encompassing both elastic and inelastic scatterings. 
The data select non-trivial, disconnected regions of this distribution, which correspond to physically distinct endothermic and exothermic regimes, in which the observed recoil is generated either close to a kinematic threshold or through the release of dark-sector mass energy. We determine the recoil spectra preferred by the observed data and show how the absence of accompanying events at lower recoil energies sharply constrains the viable parameter space. We find a kinematically tuned region of endothermic WIMP-like models at high mass for generic couplings, as well as exothermic models ranging from sub-GeV to heavy dark matter with sizable mass splittings. Elastic dark matter scatterings require instead momentum-suppressed and/or nuclear-spin-dependent non-relativistic interactions.}

\maketitle

\section{Introduction}
\label{sec:intro}
The Lux-Zeplin (LZ) collaboration recently released a search for high energy nuclear recoil (NR) events \cite{LZ:2026axp}.
LZ,  together with Xenon1t and PandaX, is the prototypical dual-phase detector for DM direct searches. One of their experimental targets is the reconstruction of NR events from the (non-relativistic) scattering of dark matter (DM) off nuclei. In \cite{LZ:2026axp} one NR event was observed with reconstructed energy $E_R(\mathrm{keV})=247\pm 23_{\rm stat.} \pm 23_{\rm sys.}$. The observation corresponds to 220 live days of data, and a fiducial volume of $\left(4.71\pm0.08\right)$ tonne, for a total exposure of $\left(2.84\pm0.05\right)$, and was selected within a specific region of interest. The experimental collaboration quoted a global significance of $2.6\sigma$ when compared to the null hypothesis, that is, SM background. Local significance reaches slightly higher values. While these result do not correspond to a discovery-level significance, it is certainly fascinating to interpret the event as arising from DM scattering off a nucleus.

From this viewpoint, the LZ event stands out with respect to expectations from the simplest WIMP DM models, which often tend to predict a larger fraction of the signal into the lower recoil-energy bins. Indeed, for non-relativistic scattering, the observed $E_R$ corresponds to a momentum transfer of $q\approx 250$ MeV, an energy at which DM starts to resolve the nucleus, thus lowering the total coherent scattering rate. This suggests that the kinematic of the scattering might be different from the usual elastic limit -- or that DM interacts with the SM with couplings that become softer and softer as the exchanged momentum diminishes, compensating the relative suppression of decoherence at such a large exchanged momentum. 

With just one event so far, every speculation or commitment to a given DM model may be unmotivated, but we ask ourselves what we can learn (globally) if we interpret the LZ event as a sign of DM.
The aim of this paper is therefore simple and direct: we derive the posterior distribution of the fundamental parameters of the dark sector allegedly responsible for the LZ event. Our model is based on the hypothesis that LZ event be described as DM, $\chi$, scattering off a nucleus $T$
\begin{equation}\label{eq:scattering}
    \chi + T \to \chi' + T\,.
\end{equation}
There are but a few parameters in play: $i)$ the DM mass, $m_\chi$; $ii)$ the mass of an hypothetical DM partner $\chi'$, $m_\chi +\delta$, where $\delta$ is the mass splitting; $iii)$ and the effective couplings $c_i$ of DM to nucleons (or more in general to SM states). We believe it is an interesting exercise to encompass all possible dark sectors by deriving  the posterior probability distribution
 \(   \mathcal{P}(M_\chi,\delta,\{c_i\})\)\,. 
With this function we can ask very general questions about the nature of the LZ event. What DM masses are preferred? What hierarchy in the spectrum of the dark sector is preferred? What are the possible coupling to the SM? We answer these questions without including strong prior beliefs -- except, of course, for the optimistic belief that the LZ event be a sign of beyond-the-SM physics -- to blindly explore the realm of possibilities. For example, we do not initially make an assumption on the sign of $\delta$, and this will give us the possibility to make a global statement on the odds of LZ to be associated more favorably to an endothermic ($\delta>0$), an exothermic ($\delta<0$), or to the elastic ($\delta=0$ and $\chi=\chi'$) process. We consider both spin-independent (SI) and spin-dependent (SD) interactions of $\chi$ with matter.

\subsection{Relation to other works}
The LZ announcement and paper release have triggered a lot of interest and a variety of theoretical interpretations, which separates into DM interpretations based on elastic and inelastic scattering, SI and SD scattering, as well as non-standard scenarios such as DM components with boosted velocity.
A large fraction of this literature has focused on inelastic DM in the endothermic regime, in which DM up-scatters into an invisible heavier state. This suppresses low-energy recoils and pushes the signal towards the upper end of the experimentally accessible window. This possibility has been investigated both in model-independent studies and in explicit realizations based on the Higgsino \cite{Freese:2026sga,Rodd:2026tyn, Du:2026guj, Yin:2026jnn} (see, however, \cite{Pospelov:2026ewn, Nguyen:2026lui, Bose:2026ndd} for a discussion on its challenges), electroweak multiplets \cite{Smirnov:2026aqk, Visinelli:2026kgt}, singlet-doublet sectors, dark photons and other portal constructions \cite{DiMauro:2026ldr,McCabe:2026crm,Yamashita:2026ump,Bandyopadhyay:2026gjw,Borah:2026zwf,Bisal:2026khf,Okada:2026eol, deLima:2026shq, Cabo-Almeida:2026uqw, Okada:2026upm, Qi:2026vyp,Kumar:2026lgi, Yuan:2026djt, Lee:2026wof,Das:2026uyy, Kotlarski:2026pep}. The opposite regime, namely the exothermic inelastic scattering, provides a qualitatively different mechanism: DM down-scatters to a lighter dark sector state (or a neutrino \cite{Lou:2026idn}). The dark sector mass splitting translates directly into the recoil energy, allowing large-energy NRs without requiring the incoming particle to populate the extreme high-velocity tail of the galactic distribution \cite{Dent:2026bji,Fan:2026hzw, Baer:2026fpy}. Interpretations based on the elastic scattering of DM have also been considered, typically relying on SD nuclear responses, momentum-suppressed interactions, or more general non-relativistic operators whose recoil spectra are harder than those of the standard SI contact interaction \cite{DiMauro:2026ldr,Unwin:2026rdp,Elahi:2026vlm}. A further class of proposals abandons the assumption that the event originates from the scattering of ordinary virialized halo DM. These include boosted dark-sector particles \cite{Alhazmi:2026efz,Kannike:2026qyl,Liang:2026coz, Heikinheimo:2026kwp}, dark-matter absorption, neutrino-induced processes \cite{Jeesun:2026vzo,Chattaraj:2026fxn}, transition-dipole and composite dark-matter scenarios \cite{Asadi:2026iot,He:2026hqz}, as well as more exotic nuclear interpretations \cite{Gu:2026vto,Aghaie:2026vsu, Lee:2026zbr}. Complementary studies have emphasized the role of the high-energy sideband, annual modulation, solar capture, and other probes in discriminating among these possibilities \cite{Rodd:2026tyn,McCabe:2026crm,DiMauro:2026dqp,Langhoff:2026ujr}.

In this work, rather than focusing one specific microscopic interpretations, or choosing the sign of the mass splitting from the start, we first formulate the problem directly in terms of the non-relativistic scattering parameters. We then infer the posterior distribution in the global $(M_\chi,\delta)$ plane, allowing elastic, endothermic and exothermic configurations to emerge from the data within a common statistical framework. This provides a model-independent way of identifying which kinematic regimes and nuclear responses are preferred before matching the interaction onto a specific ultraviolet completion.

\section{High energy nuclear recoils: kinematics and rates}
\label{sec:formule}
We start by deriving the relevant kinematic quantities and the scattering rate for the process of interest, shown in \cref{eq:scattering}: $T$ is the nuclear target\footnote{Here we assume that the nucleus does not get excited. See \cite{Gu:2026vto} a study of this case.}, with mass $M_T$, while $\chi$ and $\chi'$ are two dark sector particles, with mass difference $\delta\equiv M_{\chi'}-M_\chi$. We keep the value of $\delta$ generic, and separate the three cases of elastic ($\delta=0$), endothermic ($\delta>0$) and exothermic ($\delta<0$) when necessary.

In the laboratory frame, where the target is initially at rest, the momentum exchanged by the DM scattering is
\begin{equation}\label{eq:q2ER}
\vec q\,^2 = 2 M_T E_R\,,
\end{equation}
where $E_R$ is the nucleus recoil energy. By total energy conservation, we can write
\beq
\vec v \cdot \vec q = \frac{q^2}{2\mu_T} + \delta\,,
\eeq
where $\vec v$ is the DM velocity in the rest frame of the nucleus, and $\mu_T$ is the nucleus - DM system reduced mass. Here and in the following, we pose $q\equiv|\vec q|$ and $v\equiv |\vec v|$.
It follows that the minimum DM velocity required to have the nuclear recoil energy $E_R$ is\footnote{Let us notice that the expression in \eqref{eq:vmin} applies in the regime where both $\chi_{1,2}$ are non-relativistic. Allowing for a value of $|\delta|$ so large that $\chi_2$ become relativistic in the exothermic case, modifies the above expression to 
\beq
v_{\min}(E_R,\delta;M_\chi) = \frac{\left| |q| {\cal F} + E_R \sqrt{{\cal F}^2 + E_R(2m_T-E_R)} \right|}{E_R {\cal F} + |q|\sqrt{{\cal F}^2 + E_R(2m_T-E_R)}}\,,
\eeq
where $|q|$ is taken from \cref{eq:q2ER}, and we defined
\beq
{\cal F} = \delta +\frac{m_T}{\mu_T}E_R -E_R +\frac{\delta^2-E_R^2}{2M_\chi}\,.
\eeq
We use this more general expression in our numerical evaluations.
}
\begin{equation}\label{eq:vmin}
    v_{\rm min}(E_R,\delta)=\frac{1}{\sqrt{2M_T E_R}}\left|\frac{M_T E_R}{\mu_T}+\delta\right|.
\end{equation}
  
Imposing energy conservation we can also compute the allowed range of recoil energies, $[E^-_R,E^+_R]$. For a fixed $v$, we have
\begin{equation}\label{eq:ERbin}
\begin{split}
E_R^\pm(v) &= \frac{\mu_T^2}{2M_T}
\left(
v\pm\sqrt{v^2-\frac{2\delta}{\mu_T}}
\right)^2\,, \\ 
\Delta E_R &\equiv E_R^+ - E_R^- =\frac{2\mu_T^2}{M_T} v\sqrt{v^2-\frac{2\delta}{\mu_T}}\,, \\
\bar E_R(v) &=\frac{\mu_T^2}{M_T}\bigg(v^2-\frac{\delta}{\mu_T}\bigg)\,,
\end{split}
\end{equation}
for the edges, size and average of the energy recoil range, respectively.

The event rate per recoil energy per unit detector mass, in the laboratory frame, can then be expressed as
\begin{equation}\label{eq:rate}
    \frac{dR}{dE_R}\equiv \frac{\rho_\oplus}{M_\chi}\sum_{T}\frac{\xi_T}{M_T}
\int\limits_{v_{\rm min}(E_R,\delta)} d^3 v\, f(\vec v)\, v\,
\frac{d\sigma_T}{dE_R}(v,E_R)\,.
\end{equation}
Here $\rho_\oplus$ is the DM density at Earth, 
while $\xi_T$ is the target nuclear isotope mass fraction (we sum over all isotopes present in the LZ detector). The differential cross section of DM scattering off nuclei is $\dd\sigma_T/\dd E_R(v,E_R)$, which we discuss in greater detail in \cref{sec:CrossSection}. Finally, $f(\vec v)$ is the velocity distribution in the lab frame, obtained from the galactic distribution boosting to the Earth's frame. \footnote{We take the vacuum speed of light to be unity, unless we report values of velocities with explicit units of length over time.} When not specified otherwise, $f(\vec v)$ is the boosted image of the standard halo model (SHM) as defined in appendix \cref{app:f(v)}, where other parametrizations are also discussed.

The spectral features of $\dd R/\dd E_R$ depend upon two main effects: kinematics, through $v_{\rm min}(E_R)$ and dynamics, through $\dd\sigma/\dd E_R(v,E_R)$. In this section we want to isolate the kinematic effects, assessing what is the \emph{DM-luminosity} as a function of $\delta$, to be then convoluted with the cross-section. 

Kinematics alone already gives a bias towards particular $E_R^*$, depending on the size of the mass splitting. At very high DM velocities we expect a universal behavior for not-too-extreme values of the splitting, $|\delta| \ll \mu_T v^2$. Such a high velocity region will be mostly cut-off by the DM velocity distribution, as well as by the nuclear form factors (FFs), once the cross-section is included. On the contrary, for typical DM velocities, as well as in the low velocity band, where the DM flux is maximal, the value of $\delta$ strongly affects the DM-luminosity. For the three cases, we have
\begin{description}
    \item[Elastic] $\delta=0$. The minimum velocity scales like $v_{\rm min}\propto \sqrt{E_R}$, leading to an expected sizable DM-luminosity at small recoil energy, where direct-detection experiments have the maximal sensitivity. 
    
    \item[Endothermic] $\delta>0$. The recoil energies live in the $\Delta E_R$ range shown in \cref{eq:ERbin}. The splitting thus introduces a kinematic threshold, absent in the other cases, as $v_{\rm thr.}=\sqrt{2\delta/\mu_T}$. At threshold, the typical recoil energy is $E_R^*=\bar{E}_R(v_{\rm thr})=\mu_T|\delta| /M_T$, with a width controlled by deviation from $v_{\rm thr}$.
    
    \item [Exothermic]  $\delta<0$. In this case the minimum velocity is zero, and $E_R^*\approx E_R(0)=\mu_T|\delta| /M_T$, where the spectrum is expected to peak. 
    Differently from the endothermic case, the feature survives even at zero velocity, as for $v\to0$ we have $\Delta E_R\propto \sqrt{v^2|\delta|}$.
\end{description}
For a given $\delta$, we can fix the DM kinematics and flux, and compute the year-average of the inverse velocity as
\begin{equation}\label{eq:eta}
    \bar \eta(E_R)\equiv \int \frac{dt}{T_{\rm year}}\,\eta(E_R,t)\,,\quad\quad     \eta(E_R,t)\equiv \int_{v_{\rm min}(E_R,\delta)}d^3v\,\frac{f(v)}{v}\,.
\end{equation} 
Note that this quantity is computed independently of the interaction cross-section, but it already includes the overall $1/v^2$ dependence of the cross-section is already included in the definition of $\eta(E_R,t)$.
In \cref{app:f(v)} we discuss this quantity in some illustrative models.

\subsection{Cross sections}
\label{sec:CrossSection}
Non-relativistic scattering of DM off nuclei can be described in terms of the amplitude of non-relativistic scattering of DM off nucleons $N$, convoluted with the relevant nuclear response FF. One can build an effective Lagrangian by using Galilei-covariant objects, namely: ${\vec q}$ and ${\vec v}^\perp = {\vec v}_{\rm rel} + {\vec q}/2\mu_N$, where ${\vec v}_{\rm rel}$ is the DM - nuclear relative velocity, and the nucleon and DM spins, ${\vec S}_N$ and ${\vec S}_\chi$. With these invariants, the effective Lagrangian is
\be\label{eq:nreft}
{\mathscr L}_{\text{non-rel}}=\sum_{i,N} c_i^N \mathcal{O}_i^N\,,
\ee
where $N=p,n$ and $i$ scans over all the operators, whose complete basis has been identified in Ref.~\cite{Fitzpatrick:2012ix}. In our normalization, the Wilson coefficients have dimensions of inverse squared mass $c_i \sim 1/{\rm mass}^2$ and, since the $\mathcal \Op_i$ correspond to non-relativistic amplitudes, they are real. They may have a power-law dependence on $q^2$ (conventionally normalized to the nucleon mass squared). Since \eqref{eq:nreft} is interpreted as a non-relativistic EFT, one can interpret the coefficients as indicators of the UV physics scale that generated the interaction~\cite{Bishara:2016hek,Bishara:2017pfq,Brod:2017bsw}. 

The differential cross-section can be written as
\begin{equation}
    \frac{d\sigma_T}{dE_R}=\frac{M_T}{2\pi}\frac{1}{ v^2}\sum_{N,N'}\sum_{i,j}c^N_i c^{N'}_j F_{ij}^{NN'}(v,E_R)\,,
\end{equation}
where $F_{ij}^{NN'}$ is the form factor for the channels $ij$ of nucleons $NN'$.  We can then rewrite the rate in \cref{eq:rate} as
\begin{equation}
      \frac{dR}{dE_R}=\sum_T \xi_T \frac{dR_T}{dE_R}\,,\qquad  \frac{dR_T}{dE_R}=\frac1{2\pi}\frac{\rho_\oplus}{ M_\chi} \sum_{NN'} \sum_{ij}c_i^N c_j^{N'} \,\left\langle v^{-1}  F_{ij}^{NN'}\right\rangle(E_R)\,,
\end{equation} 
where we use the notation $\langle X \rangle \equiv \int_{v_{\rm vim}} d^3 f(v) X$.  Such an average can be non-trivial due to $v$-dependence of some of the non-relativistic operators (e.g. $\Op_{5}$). However, it is entirely ascribed to the DM-nucleon matrix element, while no $v$-dependence is expected from the nuclear form factors, which are just generic functions of $q^2$.
In most cases then  -- and in all the cases discussed in this work -- we can simply factorize the velocity average, obtaining
\be
\langle v^{-1} F_{ij}^{NN'}\rangle(E_R) = \bar{\eta}(E_R)\times F_{ij}^{NN'}(E_R).
\ee
The above equation makes the importance of $\bar{\eta}(E_R)$ in the determination of the $E_R$ spectrum of the signal manifest.

For simplicity, in our analysis we only switch on one operator at time.  We consider two representative cases, 
\begin{equation}
    \Op_1^N = \mathbf{1}_{\chi}\mathbf{1}_{N}\,,\quad \Op_4^N=\vec{S}_\chi \cdot \vec{S}_N\,,
\end{equation}
which give the rates
\begin{eqnarray}\label{eq:R1}
      \frac{dR_T}{dE_R}\bigg|_{\Op_1}&=&\frac{(c_1)^2}{2\pi}\times \frac{\rho_\oplus}{M_\chi} \times \bar\eta(E_R)  \sum_{NN'}F_{11}^{NN'}\\
    \frac{dR_T}{dE_R}\bigg|_{\Op_4}&=&\frac{(c_4)^2}{2\pi}\times \frac{\rho_\oplus}{M_\chi}\times \bar\eta(E_R) 
    \sum_{NN'}F_{44}^{NN'}\label{eq:R4}
\end{eqnarray}
Furthermore, we work in the nucleon isospin-symmetric limit. In this regime, simple explicit expressions for the nuclear FF $\sum_{NN}F_{ii}^{NN'}$ can be obtained. These are given in \cref{app:FF} for a xenon nucleus, together with all the xenon isotopes considered, and their relative fractions.

\section{Data, likelihoods and dark matter parameters estimation}
\label{sec:DataLikelihood}
In this section we construct a simplified likelihood, $\mathcal{L}_{\rm LZ}$, for the LZ results. Assuming that events in an energy recoil bin, $b_i$, are distributed as a Poisson distribution, we have 
\begin{equation}\label{eq:lkl}
    \mathcal{L}_{\rm LZ}= \prod_{b_i}\frac{e^{-\mu_i}\mu_i^{N_{i}}}{N_{i}!}=e^{-\mu_{\rm tot}}\mu_{X}\,,
\end{equation}
where $\mu_i$ is the expected number of events in the $i$-th bin, with $\mu_{\rm tot}=\sum_i\mu_i$, and $N_i$ is the observed one. Given the very low rate of expected background events across the bins, we assume that it is effectively negligible. In the last step, we simplify the sum in \cref{eq:lkl} by considering that only one event has been observed, without events at lower recoil energy. We call $\mu_X$ the expected number of events in the bin where the event was observed.
Inspired by the reported uncertainty, we take $23\sqrt{2}$~keV$\approx33$~keV-wide bins in the region where LZ has sensibility larger than 50\%, namely between $5.4$~keV and $269.9$~keV \cite{LZ:2026axp}. With this choice, the observed NR lies in $b_X=[236.4,269.4]$~keV.

\begin{figure}[t]
    \centering
    \includegraphics[width=0.48\linewidth]{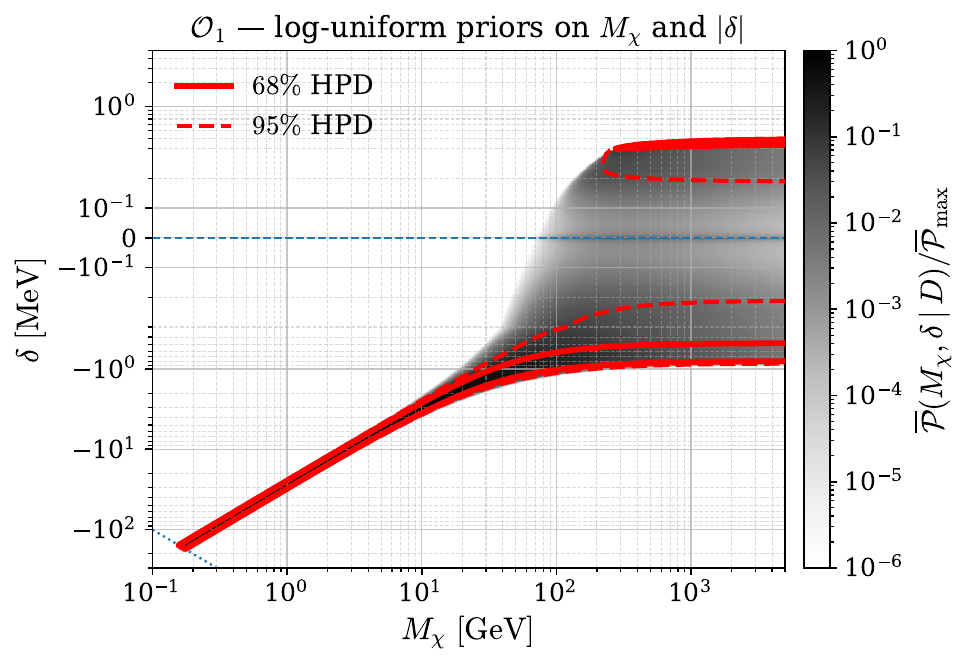}
    \includegraphics[width=0.48\linewidth]{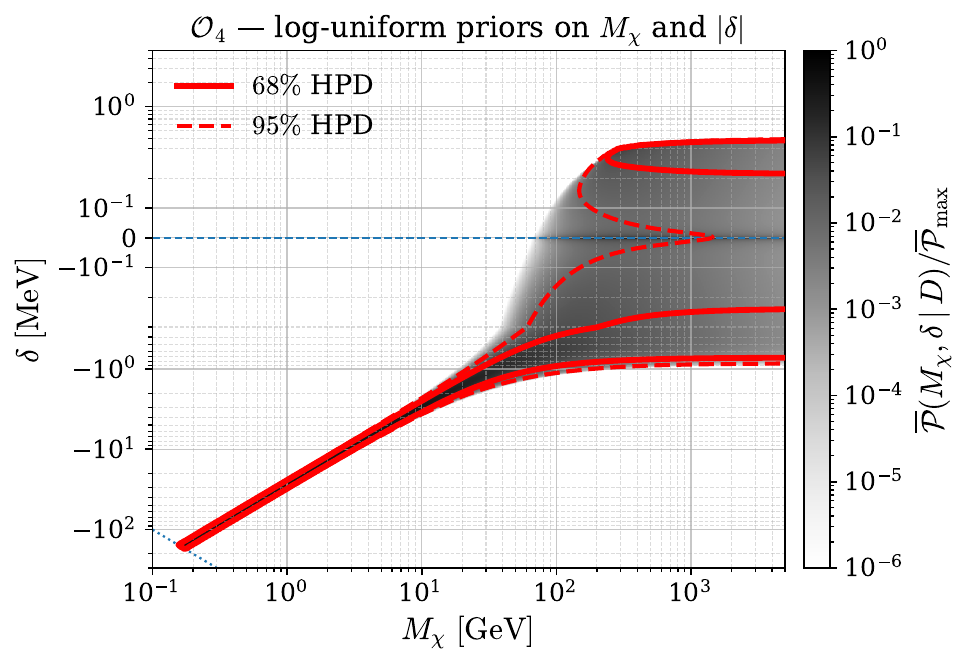}
    \caption{Posterior probability for the model $\Op_1$ (left) and $\Op_4$ (right), in the $(M_\chi,\delta)$ plane. The posterior has been marginalized over $c_1$, and for all parameters we assume a log-uniform prior (see text for details). The solid and dashes red lines show the 68\% and 95\% HPD regions, respectively. }
    \label{fig:O1_full}
\end{figure}

At this point we are in the position to infer the posterior distributions of the model's parameters $\vec p\equiv \{c_i,M_\chi,\delta\}$, and their correlations. Namely, we aim to compute $\mathcal{P}(\vec p)\propto \mathcal{L}_{\rm LZ}\times \pi(\vec p)$, where $\pi(\vec p)$ is the parameter's prior distribution.

We can further simplify the inference: since we focus on one operator at a time, it follows the event rate is simply proportional to the overall factor $c_i^2$. We can define $\mu_j \equiv c_i^2 \tilde\mu_j$, and marginalize over $c_i$ as
\beq
\overline{\mathcal{P}}(M_\chi,\delta) = \int {\cal P}(\vec p)\dd c_i\,,
\eeq
The resulting two-dimensional posterior depends on the assumed prior. For example, by taking a uniform and a log-uniform prior over $c_i$, we have
\begin{equation}
    \overline{\mathcal{P}}(M_\chi,\delta) \propto\left\{ \begin{array}{cr}
    \displaystyle  \frac{\tilde\mu_X}{\tilde\mu_{\rm tot}^{3/2}}   & \quad \text{uniform}\vspace{2mm}\\
     \displaystyle \frac{\tilde\mu_X}{\tilde\mu_{\rm tot}}    & \quad\text{log-uniform}
    \end{array} \right.
\end{equation}
Since here we would like to blindly explore the parameter space we take the log-uniform prior over $c$, weighing vastly different values of $c$ in roughly the same way. In light of the shapes and amplitudes of $\bar\eta(E_R)$ and nuclear form factors, this will allow us to cover in one analysis several kinematic configurations that have intrinsically very different rates for the same value of $c$. For the very same reason, we also use log-uniform priors for $M_\chi$ and $\delta$.

As discussed in \cref{sec:CrossSection}, the sign of $\delta$ depends on the nature of the DM model, and can in principle be fixed a priori. Before such a commitment, it is interesting to explore the global $(M_\chi,\delta)$ plane, and find the highest probability density (HPD) regions. The results for the two cases of $\Op_1$ and $\Op_4$ are shown in \cref{fig:O1_full}, where the solid and dashed red lines indicate the 68\% and 95\% HPD regions, respectively.
In both cases, the posterior probability $\overline{\mathcal{P}}(M_\chi,\delta)$ features a multi-modal behavior, with two disconnected, high-probability regions, at positive and negative values of $\delta$, whose separation is enhanced by the choice of the log-uniform prior for the Wilson coefficient. In the vicinity of the elastic case, the posterior drops, disfavoring it as a possible explanation of the excess. The drop is however different for $\Op_1$ and $\Op_4$: in the former case, the SI FF leads to a large coherent enhancement, which would give a large rate of events in the bin of interest, while for the latter the SD FF strongly suppresses the signal. A comparison of the two FFs are shown in \cref{fig:response-nuclear}. 

In the case of $\Op_1$, the elastic limit is not encompassed by the 95\% HPD region. The reason being that the corresponding nuclear FF is highly suppressed ($\sim 10^{-4}$, see \cite{DiMauro:2026dqp}) at the observed recoil-energy. In the case of $\Op_4$, instead, the FF suppresses the rate less dramatically, hence there is a marginal consistency. Note that, as we discuss below and in the next section, this consistency is enhanced by the choice of the log-uniform prior on $\delta$. For these reasons, we compute the posterior mass distribution for the endothermic and exothermic branches 
\begin{equation}
P_{\pm}= \int_{\delta\stackrel{>}{<}0} \overline{\mathcal{P}}(M_\chi,\delta) d M_\chi  d  \delta \,,
\end{equation}
as a means to compare the two and determine which is the most favorable. We also compute the maximum \emph{a posteriori} (MAP) for $m_\chi$ and $\delta$. As can be seen in \cref{tab:posterior_summary}, from the point of view of data, the global posterior (for all signs of $\delta$) penalizes the endothermic $\delta>0$ branch, as there is much more volume in the negative splitting region.

\begin{table}[t]
\centering
\caption{Global posterior sign probabilities and global MAP point for $\mathcal{O}_{1,4}$ with log-uniform priors and uniform priors on mass and mass splitting.}
\begin{tabular}{c c c c c c c}
\toprule
Operator & Prior &
$P_-$ &
$P_+$ &
$P_-/P_+$ &
$M_\chi^{\rm MAP}/{\rm GeV}$ &
$\delta^{\rm MAP}/{\rm GeV}$ \\
\midrule
$\mathcal{O}_1$ & log-uniform &
$0.8665$ & $0.1335$ & $6.49$ &
$0.1676$ & $-0.1657$ \\
$\mathcal{O}_4$ & log-uniform &
$0.7783$ & $0.2217$ & $3.51$ &
$0.1720$ & $-0.1713$ \\
\midrule
$\mathcal{O}_1$ & uniform &
$0.8245$ & $0.1755$ & $4.70$ &
$0.1676$ & $-0.1657$ \\
$\mathcal{O}_4$ & uniform &
$0.7551$ & $0.2449$ & $3.08$ &
$0.1720$ & $-0.1713$ \\
\bottomrule
\end{tabular}
\label{tab:posterior_summary}
\end{table}

\begin{figure}[t]
    \centering
    \includegraphics[width=0.48\linewidth]{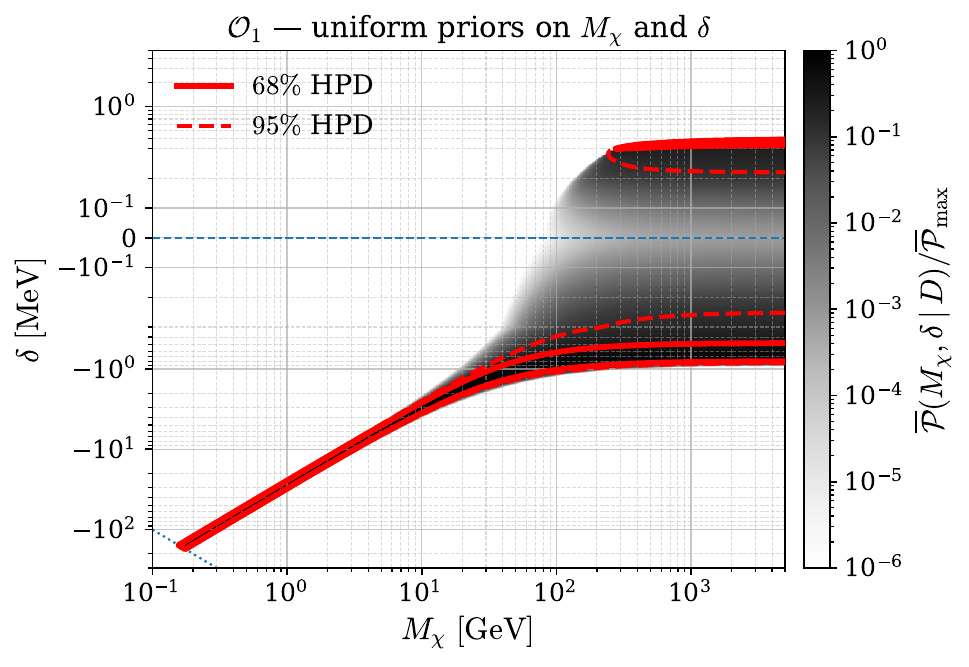}
    \includegraphics[width=0.48\linewidth]{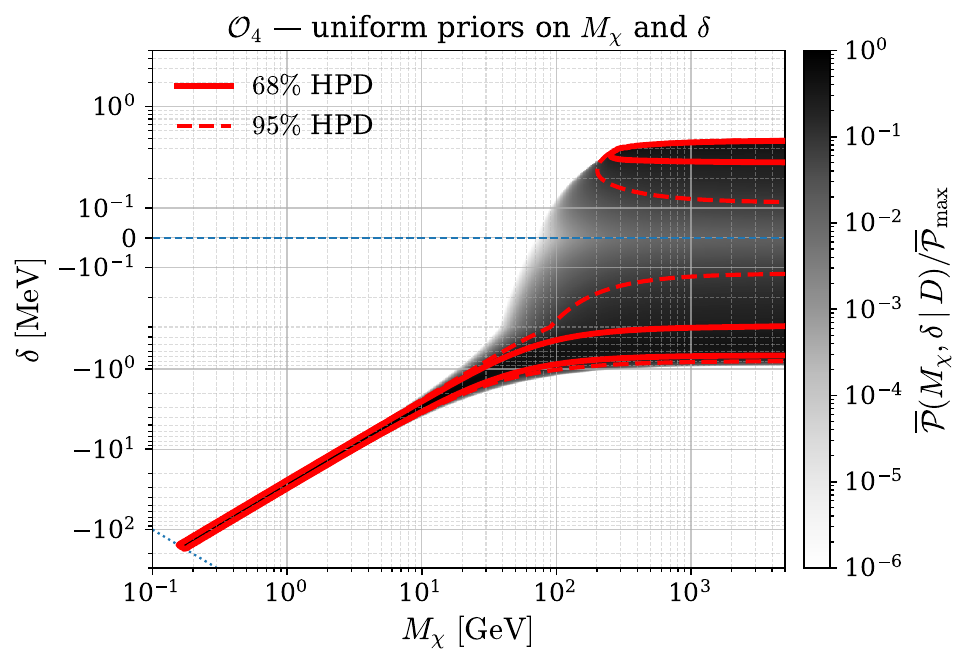}
    \caption{Posterior probability for the model $\Op_1$ (left) and $\Op_4$ (right), in the $(M_\chi,\delta)$ plane. The posterior has been marginalized over $c_1$ with a log-uniform prior, while $M_\chi$ and $\delta$ have uniform priors. The solid and dashes red lines show the 68\% and 95\% HPD regions, respectively. }
    \label{fig:uniform}
\end{figure}

Despite the log-uniform prior for $M_\chi$ and $\delta$ seems appropriate for exploring vastly different scales, it should be noted that for the case of $\delta$ it may over-emphasize the elastic scattering region. As $\delta\to 0$ the measure $d\delta/|\delta|$ leads to a bias towards the elastic scattering models. In order to check the dependence on the prior we also do the analysis with uniform priors. The results are shown in figure \ref{fig:uniform}. An immediate result of this exercise is that the compatibility at 95\% level with the $\delta=0$ is lost also in the $\Op_4$ model. On the other hand, the ratio $P_-/P_+$ and the MAP values are basically unaffected by the prior choice.

Let us summarize the main results of this section. We find a global preference for exothermic processes, with a very small mass (sub-GeV) and maximal splitting (i.e. almost massless dark-sector partner). The exothermic region shows a strong correlation between parameters that might be resolved only when focusing on specific UV realizations. The endothermic region, on the other hand, has a lower bound on the DM mass of a few hundreds of GeV. Finally, elastic processes are hardly compatible with simple models like $O_{1,4}$, and require further suppressions, e.g. by powers of $\vec{q}^2$.

\section{Interpretation and models}\label{sec:models}

From the global analysis in the previous section we see that both endothermic and exothermic scattering can reproduce the observed NR, with a global preference for the exothermic branch.
In order for purely elastic scattering to be a viable explanation of data, interactions for which the nuclear FF suppression at \(E_R\simeq 250\,\mathrm{keV}\) is not too large can be considered. Another possibility is that of considering non-relativistic operators featuring additional powers of \(q^2\). Indeed, since $q^2$ is a Galilei invariant, any operator in the non-relativistic basis can be multiplied by a power $q^{2\alpha}$ to obtain a new invariant operator. Higher exponents correspond to harder spectra, which can accommodate the non-observation of NRs of smaller energy.

\begin{figure}
    \centering
    \includegraphics[width=0.9\linewidth]{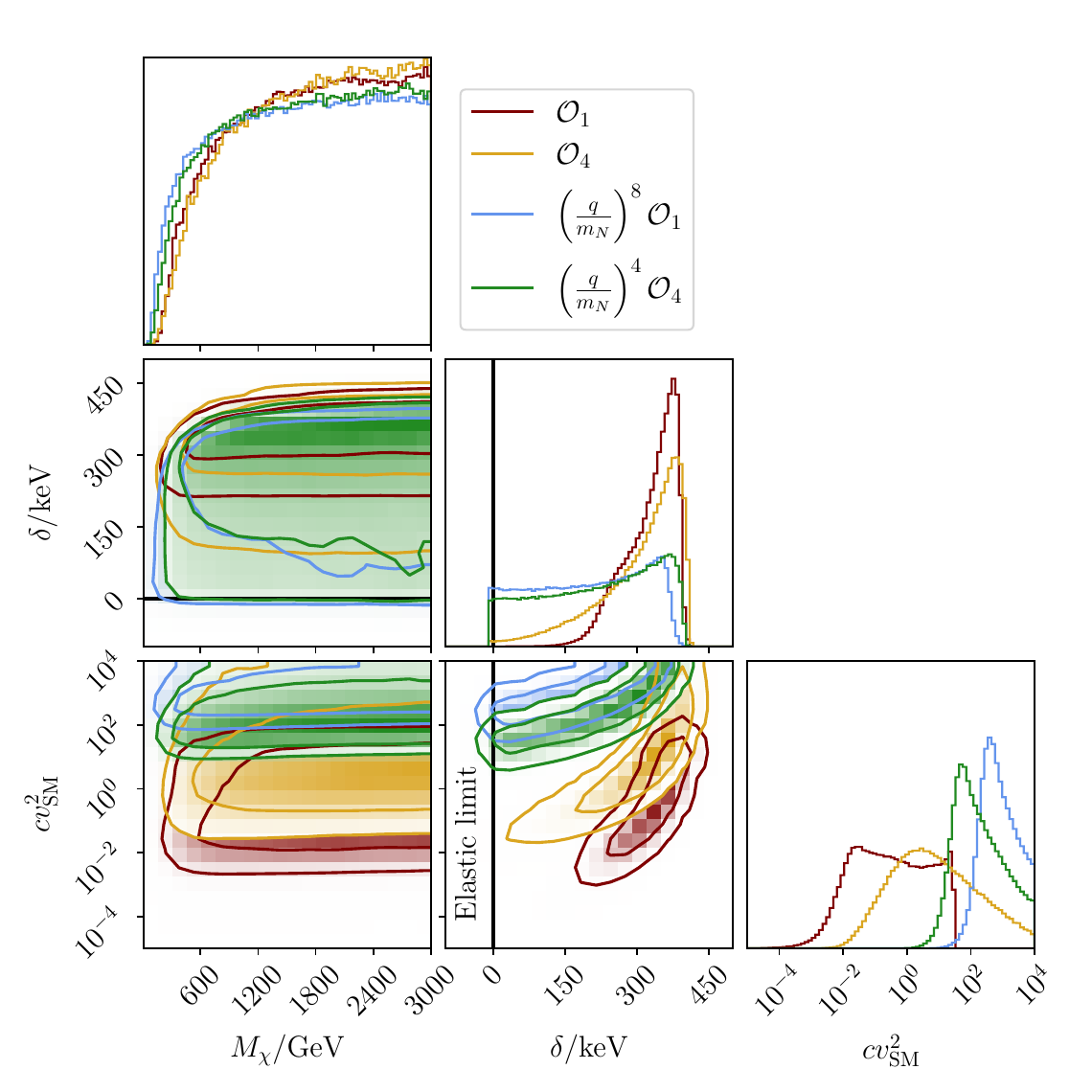}
    \caption{Posterior probability for each of the four choices of operator, $\Op_1$, $\Op_4$, $q^8 \Op_1$, $q^4 \Op_4$, when compared with the observed recoil energy, assuming uniform priors $\delta\in\left[0,1\,{\rm MeV}\right]$ and $M_\chi\in\left[10\,{\rm GeV}, 3\,{\rm TeV}\right]$. In practice, we allow the MCMC to explore small, negative values of $\delta$, both for numerical stability and to understand whether the elastic limit is allowed by a specific choice. We take a log-uniform prior on the Wilson coefficient $c\,v^2_{\rm SM}$, where $v_{\rm SM}=246$GeV.} 
    \label{fig:corner}
\end{figure}

The next question is which microscopic models can realize these possibilities.
For as far as the endothermic case is concerned, the positive sign of the splitting is natural in the sense that, if the DM is the lightest state in the dark sector, it is easily made stable. What is less natural is that the dark sector comes with mass splitting which is precisely the one needed to be seen in detectors.
On the other hand, exothermic scenarios are potentially in danger: the DM could be unstable on short cosmological time-scales, and large negative values of $\delta$ imply the existence of light states, thus potentially conflicting with observations of the cosmic micro-wave background and of Big Bang nucleosynthesis.
Finally, UV models whose non-relativistic limit corresponds to operators with extra powers of $q^2$ -- especially one operator a time -- are of difficult conception. This challenges explanations of the LZ data with elastic scatterings. When we consider instances of these operators, both in this section and in \cref{app:momentum-suppressed}, we take them as an illustration of the need of a harder spectrum rather than a believable model. 

Since, as can be appreciated from \cref{fig:O1_full}, in the negative-$\delta$ branch the posterior is extremely degenerate in the $M_\chi$-$\delta$ plane, in this section we only focus on the endothermic, positive-$\delta$ branch, which is more suited for exploring the full three-dimensional posterior in $M_\chi$, $\delta$, and $c_i$. The Wilson coefficient preferred by data gives information on the scale at which new-physics in the UV regime generates the interactions between the DM and nuclei. 
We adopt uniform priors on both $\delta\in\left[0,1\,{\rm MeV}\right]$ and $M_\chi\in\left[10\,{\rm GeV}, 3\,{\rm TeV}\right]$, while maintaining a log-uniform prior on the Wilson coefficient, in order to explore vastly different UV models. Note that these prior differ from those used in the derivation of the previous section, a fact that can have an impact on the probability of regions close to the elastic limit. Moreover, resolving the
posterior close to $\delta=0$ allows us to test whether the elastic limit
can be reached for a given choice of interaction.

We consider four types of interactions, in the language of \cite{Fitzpatrick:2012ix}
\be\label{eq:models}
\mathcal \Op_1\,,\qquad \mathcal \Op_4\,,\qquad \left(\frac{q}{m_N}\right)^8~\mathcal \Op_1\,, \qquad \left(\frac{q}{m_N}\right)^4~\mathcal{O}_4\,,
\ee
and employ a Monte Carlo Markov chain (MCMC) sampler to explore the posterior.\footnote{We make use of the python package \texttt{emcee} \cite{emcee}. The corner plots were made with the use of the package \texttt{corner} \cite{corner}.} The resulting 68\% and 95\% highest-density regions are shown in \cref{fig:corner}, each color representing a different model in \eqref{eq:models}. As expected from the kinematic threshold effects, there is a strong correlation between the Wilson coefficient $c$ and $\delta$, especially for $\Op_1$. 
Moreover, $\Op_1$ and $\Op_4$ are incompatible at 95\% with the elastic limit (black line).
In \cref{fig:O1_full}, we argued that considering a log-uniform prior knowledge of $\delta$ allowed $\Op_4$ to be marginally compatible with elastic scattering, as such a prior favors small values of $\delta$ in the posterior. We stress that in this case we consider a uniform prior knowledge on $\delta$. The momentum-suppression of the last two models in \cref{eq:models} compensates for the lack of observation of lower energy NRs, hence they are more compatible with the elastic limit. 

From the MAP of the Wilson coefficients one can estimate the scale of the UV physics generating these interactions. Let us pose
\be\label{eq:np-scale}
c_i \sim \frac{g_\ast^2}{m_\ast^2}\,,
\ee
where $g_\ast$ and $m_\ast$ are, respectively the coupling and typical mass scale of the UV theory. We report the estimates in \cref{tab:np-scale}. The cases of $\Op_1$ and $\Op_4$ fall in the ballpark of the WIMP, whereas the momentum-suppressed operators present more challenges on the model-building side as they seem to require very small couplings in order to avoid constraints on light mediators.

\begin{table}[t]
 \caption{Estimate of the UV scale from the posterior distribution of the Wilson coefficients of the non-relativistic EFT, using \eqref{eq:np-scale}.}\label{tab:np-scale}
    \centering
    \begin{tabular}{c c c c c}
    \toprule
      Operator   & $\Op_1$ & $\Op_4$ & $\left(\frac{q}{m_N}\right)^8 \Op_1$ & $\left(\frac{q}{m_N}\right)^4 \Op_4 $  \\
      \midrule
      $m_\ast/g_\ast\approx$   & $4$~TeV & $2$~TeV & $20$~GeV & $50$~GeV \\
         \bottomrule
    \end{tabular}
\end{table}


\begin{figure}[t]
    \centering
\includegraphics[width=.8\linewidth]{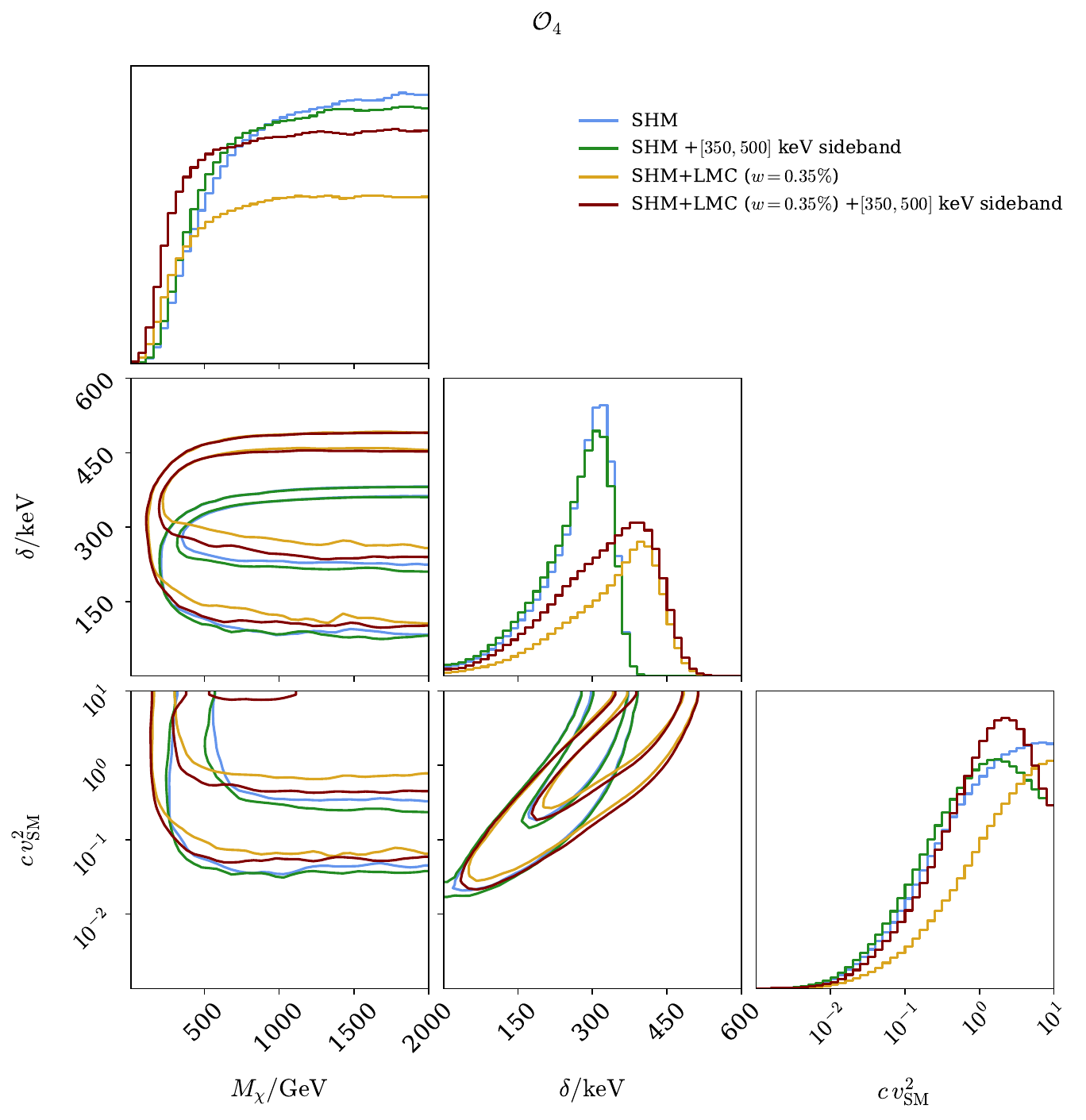}
    \caption{Posterior probability for the model $\Op_4$ with and without the high-energy `sideband'. We assume uniform priors $\delta\in\left[0,600\,{\rm keV}\right]$ and $M_\chi\in\left[10\,{\rm GeV}, 2\,{\rm TeV}\right]$.  We take a log-uniform prior on the Wilson coefficient $\log_{10}(c v^2_{\rm SM})\in [-3,1]$, where $v_{\rm SM}=246$GeV. We use two different DM velocity distributions.} 
    \label{fig:fine}
\end{figure}

\section{Conclusions and outlook}

We address the recent observation of high-energy NR at LZ under the assumption that it arose from either elastic or inelastic scattering of DM off nuclei. In the first part of the paper, we remain agnostic on the sign of the splitting between the DM and its partner, allowing for both endothermic and exothermic scattering. The idea is to understand whether the data have a preference for either case, and under what general conditions the elastic limit can be approached or reached. Allowing for different choices of priors, we find the same overall conclusion: the exothermic, $\delta<0$ case is preferred by data when compared tot he endothermic, $\delta>0$, case; the elastic limit is globally disfavored with respect to both. 

Due to the strong degeneracy of the exothermic case in the DM mass, mass splitting plane, in \cref{sec:models} we focus on the endothermic branch, which is more suited for a three-parameter inference. We considered the four models in \eqref{eq:models}, exploring the corresponding three-dimensional posterior distributions with a MCMC sampler.
We find that the elastic limit is convincingly disfavored by the data when considering the first two models. The second two models, whose operators are suppressed by extra powers of the exchanged momentum, marginally favor  the elastic limit. However, UV models reducing to these cases are particularly challenging as signaled also by the estimate of the new-physics scale to which they correspond, see \cref{tab:np-scale} and discussion.

This work can be extended in several directions. 

First, as more data become available, the inference should be performed with more datasets. However, something can be said already by noticing that, as emphasized in \cite{Rodd:2026tyn}, LZ has potential sensitivity in the high-energy \emph{sideband} region, despite the acceptance of those bins being not disclosed in \cite{LZ:2026axp}. It is interesting to redo our computation in Eq.~\eqref{eq:lkl} allowing for a high energy bin $[350,500]$ keV with zero events. In \cref{fig:fine} we show the results for the $\Op_4$ model, where we compare different datasets. We find that including the extra bin has limited effects on $(M_\chi,\delta)$, both for DM velocity distributions based on the SHM, and on SHM plus a 0.35\% boosted component denoted with LMC (see \cref{app:f(v)}). A stronger impact of the sideband is found in the posterior of the Wilson coefficients: this may pose challenges to models in which the coupling is already fixed by symmetries or power-counting. Models based on scalar operators such as $\Op_1$ will be even less sensitive due to stronger nuclear response cut-off at high $E_R$. Certainly more impactful in this regard will be the upcoming results from more exposure of LZ and from similar detectors, such as PandaX, Xenon1t and DarkSide. 

A further extension is the exploration of a larger set of models. Here, we have only focused on isospin-symmetric interaction between DM and the SM, but other choices may lead to better interpretation of the data. Moreover, a better treatment of the systematics is surely needed: to give an example, it should be possible with a MCMC to marginalize over the shape of the DM velocity distribution.

Importantly, all of the models considered meet constraints from other observations, which we did not consider in this work. These can include null results in direct searches (similar to LZ), astrophysical or cosmological probes of the cosmic micro-wave background or the Big Bang nucleo-synthesis, as well as collider searches.

Let us conclude emphasizing that the one event in the high energy bin is accompanied by many non-observations at lower energies. Those count as data, albeit null. It is because of them that -- even with a single event -- LZ can potentially discriminate among masses, mass splitting and interactions of the dark sector. This is our work's main message, \emph{a posteriori}.

{\small
\subsubsection*{Acknowledgements}

Artificial intelligence tools (ChatGPT and Claude) were used for simplifying and cross-checking our numerical codes. They were not used neither in the conception of the project nor in drawing scientific conclusions. Numerical calculations are done both with \texttt{Mathematica} and \texttt{Python} libraries. 
The work of SP and AT is supported by the Italian
Ministry of University and Research (MUR) via the PRIN 2022 project n. 20228WHTYC (CUP:I53C24002320006). MT
acknowledges support by Next Generation EU, as part of Piano Nazionale di Ripresa e Resilienza
(PNRR), Missione 4, Componente 2, Investimento 1.2 - CUP I13C25000150006. 
}

\appendix
\section{Astrophysical and nuclear inputs }
\subsection{DM velocity distribution}
\label{app:f(v)}
The DM velocity distribution used in this work, $f(\vec v)$, has been computed by boosting the SHM distribution, $f_{\rm SHM}(\vec u)$, to the Earth's frame
\begin{equation}
    f(\vec v)=f_{\rm SHM}(\vec v +\vec v_\oplus(t))\,,\,\qquad \text{with}\qquad f_{\rm SHM}(\vec u)=N \exp(-\vec u\,^2/u_0^2)\theta(u_{\rm esc.}-|\vec u|)\,,
\end{equation} 
where $N$ is a normalization such that $\int d^3u f_{\rm SHM}(\vec u)=1$.
Typical halo parameters are $u_0=220$ km/s, $u_{\rm esc.}=544$ km/s, while for $\vec{v}_\oplus(t)$  we use the values quoted in \cite{Cirelli:2024ssz}.
In order to compute the rates we need to compute
    \begin{equation}\label{eq:}
        \eta(E_R,t)\equiv \int_{ v_{\rm min}} d^3v f(\vec v) v^{-1}\,,
    \end{equation}
and then average it over one year to get $\bar\eta(E_R)$. This is the quantity that has been used in our numerical inference.

\begin{figure}
    \centering
\includegraphics[width=0.95\linewidth]{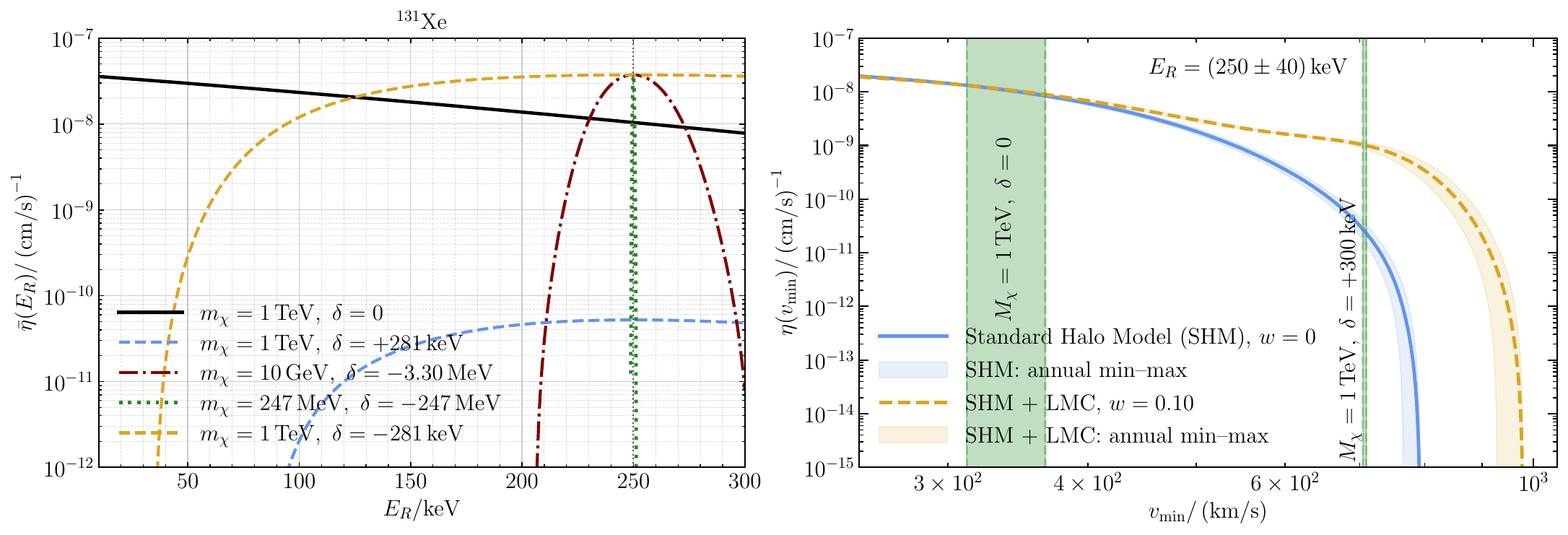}
    \caption{Left: Year averaged values of $\bar\eta$ as in eq.~\eqref{eq:eta} for a few benchmark models (for $^{131}$Xe and assuming the SHM distribution). These benchmarks include: heavy DM with both endothermic and exothermic splitting, intermediate mass DM with negative splitting and light DM with sizable negative splitting.
    Right: $\eta$ as a function of $v_{\rm min}$ for two DM velocity distributions with and without the LMC component ($w=0.1$). Two representative benchmarks of heavy DM are shown with $M_\chi=1$TeV for elastic and endothermic scattering with $\delta=300$keV. The regions correspond to the values of $v_{\rm min}$ needed to satisfy the kinematic conditions to live in the LZ high energy bin.}
    \label{fig:eta}
\end{figure}

In this appendix we quantify the impact of additional component of DM distribution with higher mean values. We adopt the same convention as in \cite{Rodd:2026tyn} adding to the SHM a boosted DM component derived from
\be
f_{\rm full}(\vec u)=(1-w)f_{\rm SHM}(\vec u) + w f_{\rm LMC}(\vec u)\,.
\ee
where $f_{\rm LMC}$ is gaussian distribution centered around $\vec u_b$ and spread $\sigma_b=100$km/s. The modulus of $u_b=570$km/s and it is at an angle $\cos\beta\approx -0.71$ with the Sun velocity. There is a cut at $200 \mathrm{km/s}=u_{\rm cut}<|\vec u-\vec u_b|$. These values are inferred from \cite{Smith-Orlik:2023kyl,Besla:2019xbx}, from studies on the Large Magellanic Cloud (LMC). For illustration purposes we show in the right panel of figure \ref{fig:eta} the impact of a very high fraction $w=0.10$ on the value of $\eta(v_{\rm min})$.

\subsection{Nuclear responses for $\Op_1$ and $\Op_4$}\label{app:FF}

In the numerical analysis, we consider a xenon detector We sum over all isotopes and we assume the following isotope mass fractions
$\xi_{128}=0.0192$,
$\xi_{129}=0.2644$,
$\xi_{130}=0.0408$,
$\xi_{131}=0.2118$,
$\xi_{132}=0.2689$,
$\xi_{134}=0.1044$ and
$\xi_{136}=0.0887$.

\begin{figure}
    \centering
    \includegraphics[width=0.45\linewidth]{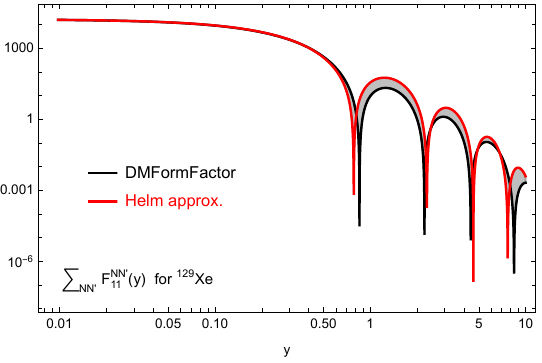}
    \includegraphics[width=0.45\linewidth]{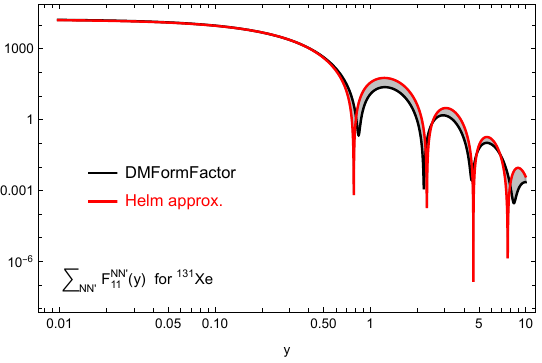}
    \includegraphics[width=0.45\linewidth]{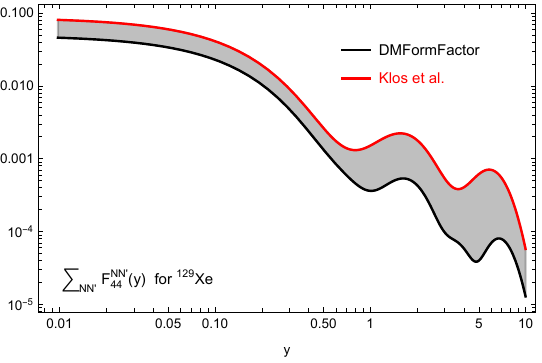}
    \includegraphics[width=0.45\linewidth]{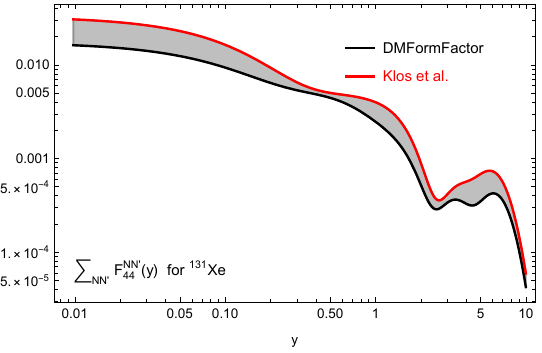}
    \caption{Nuclear responses in the isospin symmetric limit for $\Op_1$ (upper panel) and $\Op_4$ (lower panel), for two diffent isotopes of X3 (129, left) and (131, right).}
    \label{fig:response-nuclear}
\end{figure}

In this appendix we collect the expressions used in our numerical evaluations, in the isospin symmetric limit. In this limit the nuclear responses are factorized as in eq.s~\eqref{eq:R1}-\eqref{eq:R4}. We mostly employ \texttt{DMFormfactor} \cite{Fitzpatrick:2012ix,Anand:2013yka} for the evaluation of the nuclear response functions. Form factors are evaluated in the variable $y\equiv q^2 b^2/4$, where $b$ is an isotope-dependent characteristic length scale. For the SD case we always assume the DM to have spin $1/2$.

    
    \paragraph{Scalar response}
    
    For the case of $\Op_1$, using the syntax of \texttt{DMFormfactor}, we compute
    \begin{equation}
        \sum_{NN'}F_{11}^{NN'}(y)=\frac{4\times 4\pi}{2J_T+1}\texttt{ResponseNuclear}[y, 1, 0, 0]
    \end{equation}
    for each isotope. As shown in \cref{fig:response-nuclear}, it is very well approximated by the Helm approximation \cite{Helm:1956zz,Lewin:1995rx}
    \begin{equation}
        \sum_{NN'}F_{11}^{NN'}(y)=A^2 F_{\rm Helm}(y;A,R,s)\,.
    \end{equation}
    The Helm FF is defined as
$F_{\rm Helm}(q)
\equiv 
3\,\frac{j_1(qR_1)}{qR_1}\,
e^{-(qs)^2/2},
$
 where $j_1(x)$\ is the spherical Bessel function of the first kind. The parameters of the distribution are,

$R_1
=
\sqrt{
c^2+\frac{7\pi^2}{3}a^2-5s^2
}$, with $c=(1.23\,A^{1/3}-0.60){\rm fm}$, $a=0.52~{\rm fm}$, and $
s=0.9~{\rm fm}$. In terms of the variable $y$, the mapping is $b=
\sqrt{
\frac{41.467}{
45A^{-1/3}-25A^{-2/3}
}
}\ {\rm fm}$.

 \paragraph{Spin response}
 For the case of $\Op_4$, using the syntax of \texttt{DMFormfactor}, we compute
    \begin{equation}
        \sum_{NN'}F_{44}^{NN'}(y)=\frac{4\times 4\pi}{2J_T+1}\frac{1}{16}(\texttt{ResponseNuclear}[y, 2, 0, 0]+\texttt{ResponseNuclear}[y, 3, 0, 0])
    \end{equation}
    for each isotope. Given the uncertainties associated to the determination of the spin nuclear response, we also cross-checked with \cite{Klos:2013rwa,Vietze:2014vsa}. With the conventions of Tab. III of \cite{Klos:2013rwa}, where the spin response $S_{00}(u)$ is used, we have
    \begin{equation}
         \sum_{NN'}F_{44}^{NN'}(y)=\frac{\pi}{2J_T+1}S_{00}(u=2y)
    \end{equation}
    The comparison between the two approaches is shown in figure \ref{fig:response-nuclear} with overall good-agreement, see also \cite{Vietze:2014vsa} for discussions.

\section{Momentum-suppressed interactions}\label{app:momentum-suppressed}

\begin{figure}
    \centering
    \includegraphics[width=0.48\linewidth]{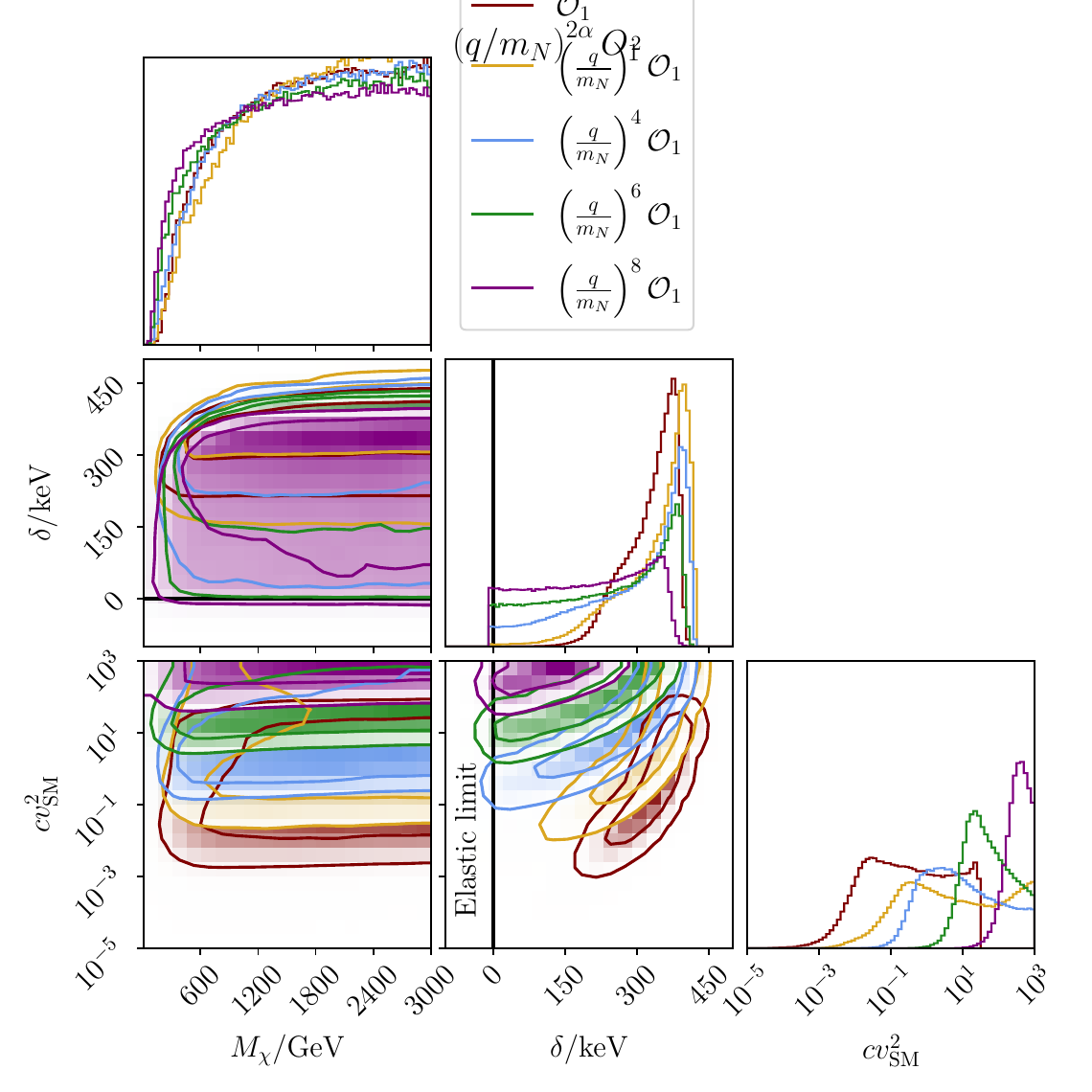}
    \includegraphics[width=0.48\linewidth]{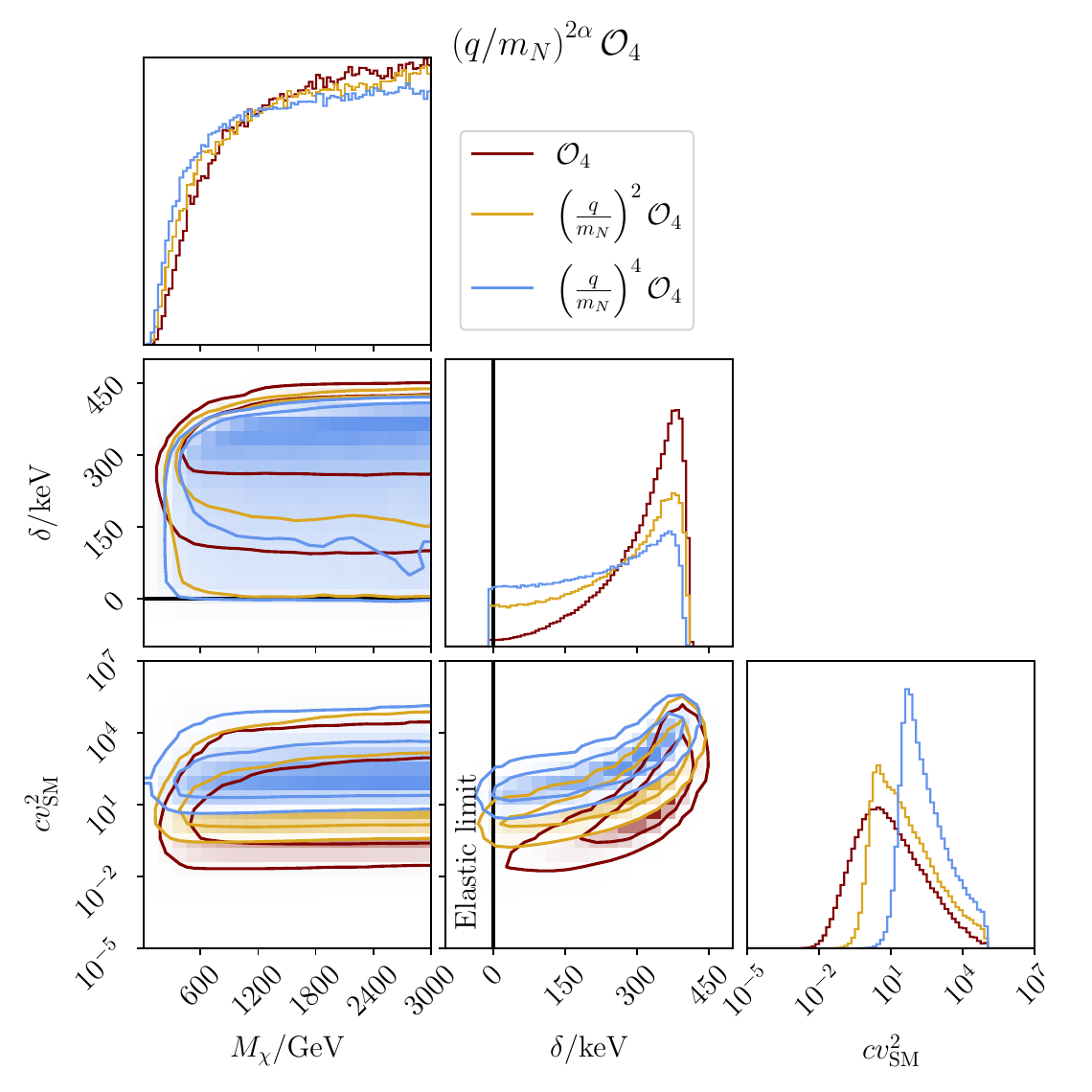}
    \caption{Posterior distributions in the case of $O_1$ (left) and $O_4$ (right), suppressed with increasing powers of $q^2$. Higher powers translate to harder spectra, which eventually allow for the elastic limit (black line) to be consistent at 95\% or 68\%.}
    \label{fig:momentum-suppressed}
\end{figure}

Models in which the DM scatters elastically off nuclei are challenged by the fact that they typically predict a larger number of low-energy NRs, in regions where no events where observed by LZ. For example, the non-relativisticc contact interaction ($\Op_1$ in the language of \cite{Fitzpatrick:2012ix}) leads to a nuclear FF peaking at low recoil energy, disfavoring the signal region by a factor of about $10^{-4}$. This is the main reason for focusing on inelastically scattering DM. Nonrelativistic interactions whose momentum-dependence favor high energy NRs may be reconciliated with the observation, but models whose non-relativistic limit goes as $q^{2\alpha} \Op_{1,4,\ldots}$ are of difficult conception. In \cref{fig:momentum-suppressed} we show the posterior distributions for $q^{2\alpha} O_1$ and $q^{2\alpha} O_4$. Higher values of $\alpha$ translate to harder spectra, which eventually allow for the elastic limit (black line) to be consistent at 95\% or 68\%. For the SD case of $O_4$, $\alpha=2$ is already sufficient to include the elastic limit at 68\%, whereas for $O_1$, on account of the strong suppression from the nuclear FF, $\alpha\gtrsim6$ is needed. 

\bibliographystyle{JHEP}
\bibliography{biblio}

\end{document}